\documentclass{aa}
\usepackage{graphicx}
\usepackage{txfonts}
\usepackage{lscape}
\usepackage{natbib}
\bibpunct{(}{)}{;}{a}{}{,}
\def\xmm{{\it XMM-Newton}}
\def\exo{EXO\,0748--676}
\def\kms{km\,s$^{-1}$}
\def\gtsim{\raisebox{-.5ex}{$\;\stackrel{>}{\sim}\;$}}

\begin{document}
   \title{Properties of the ionised plasma in the vicinity of the neutron-star X-ray binary \exo}


   \author{J.C.A. van Peet
          \inst{1}
          \and
          E. Costantini
          \inst{1}
          \and
          M. M\'{e}ndez
          \inst{2}
          \and
          F.B.S. Paerels
          \inst{3}
          \and
          J. Cottam
          \inst{4}
          }

   \offprints{E.~Costantini}

   \institute{SRON, Netherlands Institute for Space Research,
              Sorbonnelaan 2, 3584 CA Utrecht, The Netherlands
              \email{e.costantini@sron.nl}
              \and
		Kapteyn Astronomical Institute
University of Groningen
Postbus 800
9700 AV Groningen, The Netherlands
             \and
              Columbia Astrophysics Laboratory, Department of Astronomy,
              Columbia University, 550 West 120th Street, New York, NY 10027
              \and
              Exploration of the universe Division, Code 660, NASA/GSFC,
              Greenbelt, MD 20771.
             }

   \date{Received / Accepted}
\authorrunning{J.~van Peet et al.}
\titlerunning{\xmm\ observation of \exo}


  \abstract
   {}
   {We present the spectral analysis of a large set of
   XMM-\textit{Newton} observations of \exo, a bright
   dipping low-mass X-ray binary. In particular, we focus on
   the dipping phenomenon as a result of changes in the properties of the ionised gas close to the source.
   }
   {We analysed the high-resolution spectra collected with the reflection grating spectrometer on board \xmm. We studied 
   dipping and persistent spectra separately. We used the Epic data to
   constrain the broad-band continuum. We explored two simple geometrical scenarios
   for which we derived physical quantities of the absorbing material
   like the density, size, and mass.
   }
   {We find that the continuum is absorbed by a neutral gas, and by both a collisionally (temperature $T\sim70$\,eV) 
   and photoionised (ionisation parameter log$\xi\sim2.5$) absorbers. 
   Emission lines from OVII and OVIII are also detected. This is the first time that evidence of a
      collisionally ionised absorber has been found in a low-mass X-ray binary.
      The collisionally ionised absorber may be in the form of dense ($n>10^{14}$\,cm$^{-3}$) 
      filaments, located at a distance $r\gtsim10^{11}$\,cm.   
     During dips, 
      the photoionised absorber significantly increases its column density (factor 2--4) while becoming less ionised. 
      This strengthens the idea
      that the colder material of the accretion stream impinging the disc 
      is passing on our line of sight during dips. In this scenario, we find that 
      the distance from the neutron star to the impact
      region ($\sim 5\times10^{10}$\,cm) is similar to the size of the neutron star's Roche lobe. 
      The gas observed during the persistent state may have a flattened geometry. 
      Finally, we explore the possibility of the existence of material forming an initial, hotter portion 
      of a circumbinary disc.  
   }
   {}

   \keywords{Atomic processes -- Stars: binaries -- X-rays: binaries -- X-rays: individuals: EXO~0748-676}

   \maketitle
%

\section{Introduction}
Low-mass X-ray binaries (LMXB) show different types of variability in their light curve,
including X-ray bursts, eclipses, and characteristic obscuring events
called dips. A dip shows up in the light curve as a decrease
in count-rate, which can be as deep as an eclipse, but with a much less clear start and end.
Dips are probably caused by the line of sight intercepting the impact region of the accretion
stream on the accretion disc \citep{1982ApJ...253L..61W}, and are therefore seen in systems with
inclinations in the range of $60^{\circ}-80^{\circ}$ \citep{1987A&A...178..137F}.\\
Two different models have been used to explain the dipping emission.
In the first model \citep[e.g.][]{1986ApJ...308..199P} two components
(both power laws with an exponential cut off) are used to model the dipping emission.
The first component is absorbed and the second one is not. This approach
is sometimes referred to as the ``absorbed + unabsorbed approach''.\\
The second model, which consists of two types of emitting components, was proposed by
\cite{1998ApJ...504..516C}. The first component is point-like blackbody emission from the
neutron star itself, and the second component is extended Comptonised emission from the
accretion disc corona. During dipping intervals, the accretion disc corona is progressively
covered, hence the name ``progressive covering approach''.\\
Recently, \cite{2005A&A...436..195B} have proposed a different approach to modelling the dipping
spectra in the LMXB 4U 1323--62. They find that the spectral changes during dipping
emission can be modelled by varying the properties of a photoionised absorber (i.e. column density and ionisation parameter). No absorbed + unabsorbed
components or progressive and partial covering of an extended corona is necessary and
absorption lines and the continuum are modelled self consistently. \cite{2006A&A...445..179D} show
that this approach can be used on a number of other X-ray binaries exhibiting dipping behaviour,
among which also EXO 0748--676.

EXO 0748--676 is an LMXB that shows all the variability mentioned above: bursts, eclipses, and dips.
It was discovered in 1986 \citep{1986ApJ...308..199P, 1986ApJ...308..213G}
and has been studied regularly since its discovery \citep{1991ApJ...366..253P, 1995ApJ...438..385H,
1997ApJ...486.1000H, 1997ApJ...480L..21T, 1998ApJ...504..516C,
2005A&A...429..291S, 2005ApJ...632.1099W} especially since
the launch of the high resolution X-ray observatories XMM-\textit{Newton} \citep{2001A&A...365L.282B,
2001A&A...365L.277C, 2002Natur.420...51C, 2003A&A...412..799H, 2006A&A...445..179D},
and \textit{Chandra} \citep{2003ApJ...590..432J}.

\cite{1986ApJ...308..199P} derived the basic properties for EXO 0748--676.
The LMXB has a period of 3.82 hours as measured by X-ray eclipse timings,
and the eclipses themselves have a duration of 8.3 minutes. Assuming a compact
object of $1.4 M_{\odot}$, \cite{1986ApJ...308..199P} infer the mass of the companion
to be between $0.085 M_{\odot}$ and $0.45 M_{\odot}$ and the system inclination
between $75^{\circ}$ and $82^{\circ}$. Recently a strong X-ray burst was observed
\citep{2005ApJ...632.1099W} from which a distance to EXO 0748--676 was derived
that depends on whether the burst is hydrogen dominated
($5.9\pm0.9 \mathrm{\,kpc}$) or helium dominated ($7.7\pm0.9 \mathrm{\,kpc}$).
In this paper we use the average value of $6.8 \mathrm{\,kpc}$.

We study here the ionisation state of the accretion disc around the neutron star of 
EXO 0748--676 in order to get a better understanding of the geometry, dynamics and
ion stratification of the disc. We use all RGS data available in the wavelength range 7--35\AA.

In Section \ref{sec: data analysis} we describe the observations and the data analysis, in Section
\ref{sec: results} we present the results of the spectral fitting. In section \ref{sec: discussion}
we discuss the results and derive physical parameters of the system and in Section
\ref{sec: conclusions} we present our conclusions.

\section{Data analysis}\label{sec: data analysis}
The data were taken from the XMM-\textit{Newton} archives and were acquired
with the Epic PN \citep{2001A&A...365L..18S}, Epic MOS \citep{2001A&A...365L..27T} and RGS
\citep{2001A&A...365L...7D} instruments.
For the analysis of each data set we require first the availability of 
the RGS and then of either the PN or MOS cameras (Table~\ref{tab: observations}).
We processed the data with version 6.5.0 of the XMM-\textit{Newton} Science Analysis System (SAS). For the Epic analysis,
 the background was extracted 
from a circular region in the same field of view of the source. From this region we produce
a light curve from which we removed any background flares. Only four observations were 
totally discarded because of a persistently flaring background.
To obtain the source light curve
and spectrum we use an annular extraction region ($R_{in} \approx 8''$ and $R_{out} \approx 51.5''$)
to prevent pile-up. We divide the source radiation into hard ($5 \leq \mathrm{E (keV)} \leq 10$)
and soft ($0.3 \leq \mathrm{E (keV)} \leq 5$) X-ray emission. We use the hard X-ray light curve
to obtain the starting and ending times for the bursts and the eclipses, since this energy
range is marginally affected by
dipping \citep{2006A&A...445..179D}. EXO 0748--676 occasionally shows two, or even three,
 bursts in a row \citep{2007A&A...465..559B}, with an inter-burst time that is much shorter than between two
normal bursts. We count these double and triple bursts as one, since we are only interested
in the starting and ending times of the total interval containing bursts.\\
We also use the PN data to obtain the Good Time Intervals (GTI) for
the dipping and eclipsing emission (we will call persistent emission everything that is not dip, burst or eclipse). 
These GTIs were subsequently used to extract
the RGS spectra. If, for a certain observation, PN (or MOS) 
were not switched on at the same time as RGS, we only use the overlapping intervals.\\
Next, we divide the hard X-ray light curve by the soft
X-ray light curve. We use the resulting hardness ratio curve to distinguish between dipping and
persistent emission: a dip interval shows up as an increase in the hardness ratio. We use the average
of a representative portion of the hardness ratio curve that clearly belongs to the persistent
emission to select dipping and persistent emission: values higher than 2 times the
average persistent hardness ratio were selected as dipping emission.\\
Upon close inspection of the hardness ratio curves we found that
it was not possible to distinguish between persistent and dipping emission in the way
that was described above for the observations from orbits
0692 and 0708. The reason for this is that the hardness ratio curve for these observations
does not show clear separate periods of dipping and persistent emission as compared to a normal case.
For this reason, even if these two data sets satisfied our first selection criteria, we did not analyse them any further.\\
We process the RGS data with the SAS task ``rgsproc''. After we check the source coordinates
and correct them if necessary, we create a spectrum of the RGS.\\   
We apply an effective area correction to the RGS data processed with SAS version 6.5.0 
(private communication of J. Vink)\footnote{
more information can be found on the website of the XMM-\textit{Newton} users
group: http://xmm.vilspa.esa.es/external/xmm\_user\_support/\\usersgroup/20060518/rgs\_calib\_eff.pdf).
This correction has been incorporated into subsequent SAS releases.}.
We fit the area-corrected spectra with the X-ray and UV fitting package SPEX\footnote{
http://www.sron.nl/divisions/hea/spex/} \citep{1996uxsa.conf..411K}. 
The best fit was found by minimisation of the $\chi^2$. Errors are at 1$\sigma$ unless
otherwise stated. 

\subsection{The data set}\label{sec: data set}
In table \ref{tab: observations} we list the observations from the XMM-\textit{Newton}
archive used in this paper.
\begin{table*}
   \caption{XMM observation summary for EXO 0748--676.}
   \label{tab: observations}
   \centering
   \renewcommand{\footnoterule}{}  
   \begin{tabular}{l c c c c c c c c}
   \hline \hline
   orbit & obs. ID & instruments & obs. date & \multicolumn{3}{c}{observing time (ks)} & group  & spectrum\\
         &         &           & (yymmdd)  & total & per & dip                       &        & mos/pn  \\
   \hline
   0055 & 0122310301 & mos/rgs    & 00-03-28    & 61.0  & 11.9 &  0.9 & A   &  -  \\
   0067 & 0123500101 & mos/pn/rgs & 00-04-21    & 62.1  & 16.9 & 14.5 & A   & yes \\
   0212 & 0134561101 & mos/pn/rgs & 01-02-03    & 8.4   &  0.0 &  3.8 & A   &  -  \\
   0212 & 0134561201 & mos/pn/rgs & 01-02-04    & 6.6   &  0.3 &  1.7 & A   &  -  \\
   0212 & 0134561301 & mos/pn/rgs & 01-02-04    & 6.9   &  1.6 &  1.3 & A   &  -  \\
   0212 & 0134561401 & mos/pn/rgs & 01-02-04    & 6.6   &  1.5 &  1.4 & A   &  -  \\
   0212 & 0134561501 & mos/pn/rgs & 01-02-04    & 6.6   &  1.2 &  1.3 & A   &  -  \\
   \hline
   0338 & 0134562101 & mos/rgs    & 01-10-13    & 8.7   &  4.7 &  2.1 & B   &  -  \\
   0338 & 0134562201 & mos/rgs    & 01-10-13    & 6.9   &  4.8 &  1.3 & B   &  -  \\
   0338 & 0134562301 & mos/rgs    & 01-10-13    & 6.9   &  6.2 &  0.0 & B   &  -  \\
   0338 & 0134562401 & mos/rgs    & 01-10-13    & 6.9   &  2.8 &  2.3 & B   &  -  \\
   0338 & 0134562501 & mos/rgs    & 01-10-13/14 & 6.9   &  2.5 &  2.4 & B   & yes \\
   \hline
   0693 & 0160760201 & mos/pn/rgs & 03-09-21/22 & 93.4  & 41.5 & 23.1 & C2 &  -  \\
   0694 & 0160760301 & mos/pn/rgs & 03-09-23/24 & 108.3 & 47.8 & 29.6 & C3 & yes \\
   0695 & 0160760401 & mos/pn/rgs & 03-09-25/26 & 82.0  & 29.4 & 26.6 & C3 &  -  \\
   0710 & 0160760801 & mos/pn/rgs & 03-10-25/26 & 67.9  & 21.8 & 18.1 & C1 & yes \\
   0719 & 0160761301 & mos/pn/rgs & 03-11-12/13 & 94.7  & 31.3 & 35.3 & C2 & yes \\
   \hline
   \end{tabular}
\end{table*}
In the first column the orbit in which the observation was made is listed.
The second column shows the observation identifier and the third column shows which instruments
were used to observe the source. The ``obs. date'' column gives the date on which the observation
was made. The label ``observing time'' gives the time subdivided in ``total'', persistent (``per'') and dipping (``dip'') times.
Differences are caused mainly by the fact that eclipses and bursts are left out.\\
We grouped the observations, when
possible, (column ``group'' in table \ref{tab: observations}) using the SAS-task ``rgscombine''.
The first criterion for combining separate spectra is to look whether the observations
were close in time. Secondly, we fit a continuum model including absorption (as defined in the next sections) 
to the data and calculate the errors
of the model parameters. If the parameters of the separate observations are consistent
within the errors, we combine the observations.\\
We combine the observations from orbits 0055, 0067 and 0212 into group A, 
the observations from orbit 0338 into group B and the
observations from orbits 06XX/07XX in three groups. In particular, we kept the
observation from orbit 0710 as a separate group C1. We then combined the observations from orbits 0693 and 0719
into group C2, those from orbits 0694 and 0695 into group C3 (Table \ref{tab: observations}).

We use a single observation from each of the groups defined in Table \ref{tab: observations}
to create a spectrum of the PN (or MOS, if no PN is present) data and we fit both the RGS
spectra and the PN spectrum of a group simultaneously to obtain the best fit. The observation
we use to obtain the PN spectra is indicated in the last column of table \ref{tab: observations}.
There are two reasons to use PN data. The first is to constrain the power-law photon index
in the wavelength range of the RGS instrument. The second is that PN data may offer us further constrains on a possible highly ionised
absorber, whose features would show up in the 6--7\,keV band \citep[e.g.][]{2005A&A...436..195B}.

\begin{figure*}
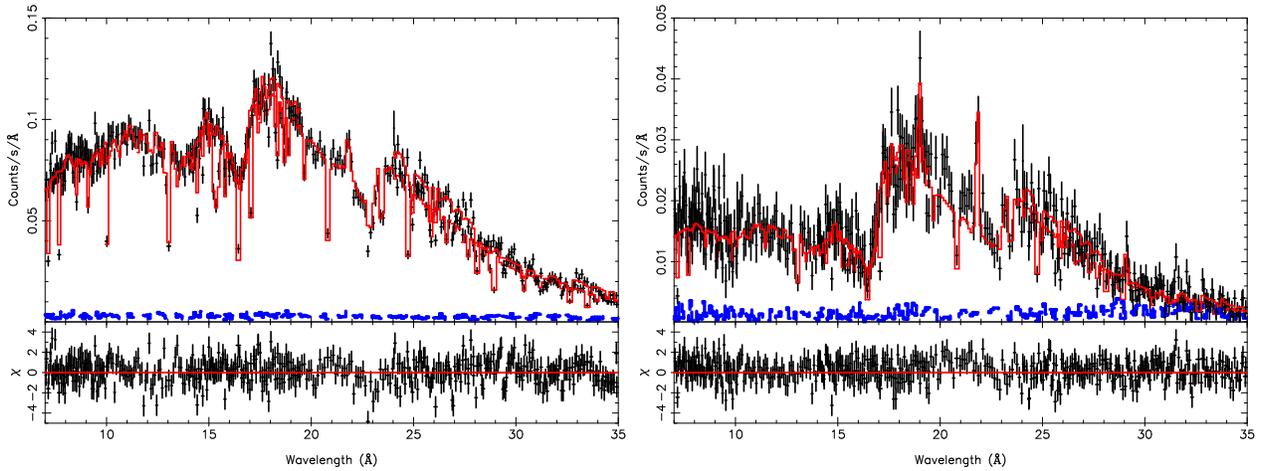

   \centering
   \resizebox{0.45\hsize}{!}{\includegraphics[angle=270]{f1_a.ps}}
   \resizebox{0.45\hsize}{!}{\includegraphics[angle=270]{f1_b.ps}}
   \caption{Two representative spectra of RGS1 and RGS2,
            taken from obsid 0160760201. The left plot shows the persistent spectrum
            and the right plot shows the dipping spectrum. In both plots the upper panel
            shows the data (points) and model (solid line) in Counts s$^{-1}$ \AA$^{-1}$. 
	    The dotted line at the bottom is the background. 
	    The lower panel shows the model residuals ((data-model)/error). The data have been rebinned for clarity.}
	    \label{fig: spectra}
\end{figure*}

\subsection{Spectral modelling}\label{sec: results, spectra description}
Two representative RGS spectra of persistent and dipping emission are plotted in Fig. \ref{fig: spectra}.
In both spectra we see a number of
interesting absorption features. At 23.1 \AA\ we see an absorption edge that
is produced by neutral oxygen in the interstellar medium. We also see two other absorption
edges, at 14.2 \AA\ and 16.8 \AA\ due to OVIII and OVII respectively. We can also 
clearly discern two emission lines, OVII at 21.8 \AA\ and OVIII at 19 \AA, the latter being more evident in the dipping
spectrum. 

As a starting point, we modelled the RGS spectra following \cite{2006A&A...445..179D}, 
who analysed the lower resolution data
of Epic-PN (orbit 0719). In particular their absorption model consisted of the combination of a neutral absorber
and one photoionised absorber. The continuum was modelled by a emission from a black body and a powerlaw. 
In our case, applying the same model
to the RGS wavelength range of 7--35 \AA\ does not yield a satisfactory fit.
In particular, the model is not capable of fitting the OVII and OVIII edges simultaneously, 
strongly suggesting the presence
of a second, ionised, absorbing component. 

\begin{figure}
   \centering
\includegraphics[height=13cm,width=9cm]{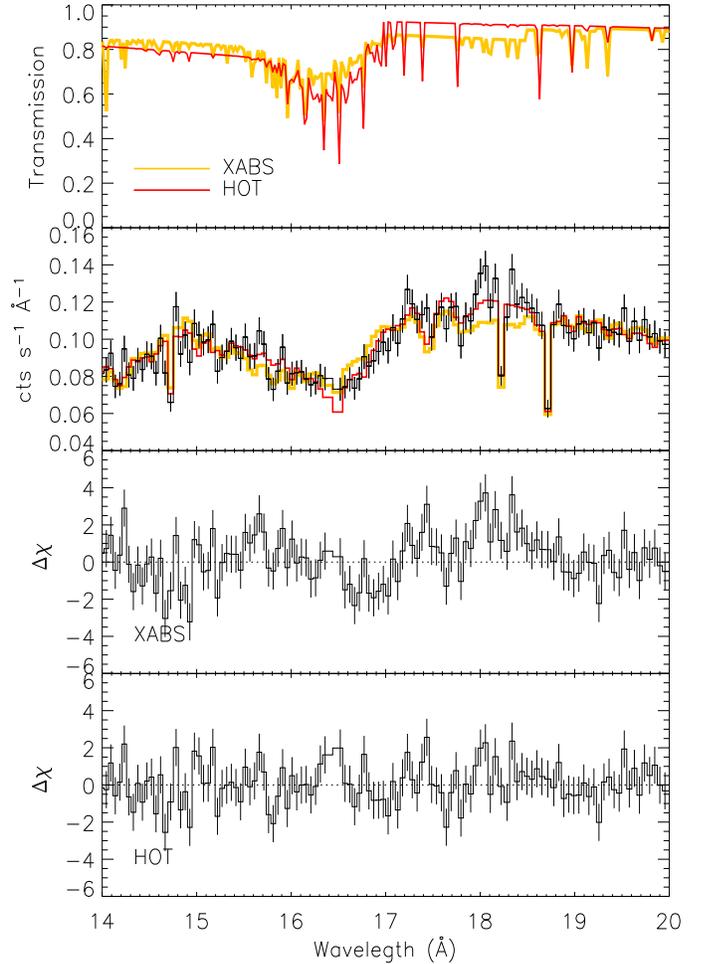}
\caption{\label{f:hot_xabs}
Comparison between the collisionally-ionised gas model ($hot$ model, dark line) and a photoionised gas 
($xabs$ model, light line) in the OVII edge-iron UTA wavelength region. This is a zoom of Fig.~\ref{fig: spectra}, left
panel. From top to bottom, first panel: theoretical transmission curve. Second panel: data points together with the $hot$ and 
$xabs$ models. The $xabs$ model fails to fit the data due to extra absorption by OV-OVI in the 17.5--18\,\AA\ 
region and in the iron UTA region. Third and fourth panels: residuals in terms of $\sigma$ to the $xabs$ and $hot$
models, respectively.}
\end{figure}

We therefore departed from this starting model to fit the high-quality, high-resolution RGS data. 
For clarity, in the following the modelling is referred to the data set shown in Fig.~\ref{fig: spectra} (left panel).
In addition to the neutral absorber and one ionised absorber (with log$\xi\sim2.5$, where $\xi$ is defined as the
ratio between the ionising luminosity and the product between the gas density and its squared distance from
the central source: $\xi=L/nr^2$) we included first
an additional photoionised absorber 
(model {\it xabs} in SPEX). This did not yield a satisfactory fit ($\chi^2/\mathrm{dof} = 2276.31/1188=1.92$, for
the persistent emission data. The column density of this additional
component is $\sim1.8\times10^{21}$\,cm$^{-2}$ with ionisation parameter log$\xi\sim0.8$. 
The OVII edge region,
close to the iron unresolved transition array (UTA), between 14 and 20 \AA,  is where the model fails the 
most in fitting the data (Fig.~\ref{f:hot_xabs}). Next we tried a more complex absorption model, which mimics a continuous distribution of
column densities as a function of the ionisation parameter (model {\it warm} in SPEX). The {\it warm} model 
did not improve the fit significantly with respect to a single-ionisation-parameter gas component.\\
The best fit is reached when a collisionally ionised gas component  is added to the model 
(model {\it hot} in SPEX) with a $\chi^2/\mathrm{dof}= 1631/1188=1.37$ for this specific data set. The gas
temperature is $\sim75$\,eV and the gas column density is $\sim1.5\times10^{21}$\,cm$^{-2}$.     
In Fig.~\ref{f:hot_xabs} we compare the two interpretations in the 14--20 \AA\ region where the OVII edge and the iron UTA
creates a broad trough in the data. In the upper panel the transmission due to the 
photoionised (light line) and
the collisionally ionised gas (dark line) are shown. In the next panel the data and the models, convolved by the instrument
resolution, are shown (with the same colour labelling as above). 
It is evident that the photoionised absorber model causes a
too shallow absorption at 16.6-17 \AA. At the same time this model severely underestimates 
the data at 17.5--18\,\AA, where
absorption by OVI and OVII takes place. On the contrary, the collisionally ionised absorption 
model provides a good fit to
the data. The reason is that given a temperature, a collisionally ionised gas ionic column density distribution strongly peaks around a single ion
(OVII in this case), while for a photoionised gas, a range of ions around a main ion is allowed. This produces the
unwanted OV and OVI absorption in the {\it xabs} model in Fig.\,\ref{f:hot_xabs}.  

The parameters for the absorption components included in the final fit are listed in Table \ref{tab: par values}. 
The underlying emission is produced by a power law,
a blackbody and two Gaussian to model the OVII triplet and OVIII Ly$\alpha$ emission lines (Table \ref{tab: par values}). 
The Gaussian profiles FWHM ranges from 0.25 to 0.5 \AA\ for OVII and from 0.05 to 
0.35 \AA\ for OVIII, respectively. 
The main component of the OVII triplet line arises around the wavelength of the
intercombination line at 21.8\AA. There is no evidence for the forbidden line at 22.1\,\AA.
The second emission line in the fit is the OVIII Ly$\alpha$ line, at
18.96 \AA. The value of the reduced $\chi^2$, reported in Table~\ref{tab: par values}, reflects the cross calibration uncertainties between RGS and PN that can be up 
to 10--15\%. 
However we are interested in narrow features located in a spectral range where instruments are well calibrated. We verified that a slight
change of the continuum shape does not affect our results.\\

\begin{table*}
\caption{Parameter values and their 1-$\sigma$ errors for all data groups.}
\label{tab: par values}
\begin{center}
\renewcommand\tabcolsep{0.15cm}
\begin{tabular}{lll|ll|ll|ll|ll}
\hline\hline
          & \multicolumn{2}{c}{group A}   &\multicolumn{2}{c}{group B}
          & \multicolumn{2}{c}{group C1} &\multicolumn{2}{c}{group C2} &\multicolumn{2}{c}{group C3}\\
parameter & pers & dip & pers & dip & pers & dip & pers & dip & pers & dip\\
\hline
\textbf{neutral abs.} & & & & & & & & & &\\
   $N_{\rm H}^a$ 
   & $1.28\pm0.05$        & $3.81\pm0.8$ & $1.10 \pm 0.03$        & $2.21\pm 0.17$
   & $1.11 \pm 0.05$        & $2.36 \pm 0.13$        & $0.82\pm0.02$ & $1.48\pm0.03$
   & $0.84\pm0.20$ & $1.25 \pm 0.03$\\
\textbf{CI} & & & & & & & & & & \\
   $N_{\rm H}^a$ 
   & $4.05\pm0.25$ & $1.8\pm1.6$ & $0.95 \pm 0.10$        & $5.28\pm0.56$
   & $3.28_{-0.40}^{+0.18}$ & $5.9\pm0.6$ & $1.95 \pm 0.07$        & $4.1\pm0.1$
   & $1.09 \pm 0.08$        & $3.42\pm0.13$\\
   $T^c$ 
   & $66\pm3$ & $<80$ & $61\pm4$ & $63\pm5$
   & $70\pm3$ & $70\pm7$ & $73\pm2$ & $69\pm1$
   & $80\pm4$ & $65\pm2$\\
   $\sigma_{v}^b$
   & $<12.27$ & $ 0. fixed $ & $<4.83$ & $123\pm30$
   & $<7.6$ & $<11.1$ & $<4.7$ & $<11.2$
   & $<3.57$ & $<15.91$ \\
\textbf{PI} & & & & & & & & & & \\
   $N_{\rm H}^a$ 
   & $41\pm4$ & $160_{-60}^{+10}$ & $37\pm4$ & $74\pm11$
   & $46\pm4$ & $140 \pm 10$      & $39\pm4$ & $80\pm3$
   & $19\pm2$ & $51\pm2$\\
   $\log{\xi} $
   & $2.35\pm0.04$ & $2.24\pm0.04$ & $2.56\pm0.04$ & $2.36\pm0.05$
   & $2.40\pm0.04$ & $2.39\pm0.03$ & $2.63\pm0.04$ & $2.38\pm 0.01$
   & $2.51\pm0.04$ & $2.33 \pm 0.02$\\
   $\sigma_{v}^b$
   & $<8.7$ & $<6.1$ & $<32.5$ & $21\pm9$
   & $<48$ & $11_{-8}^{+5}$ & $17\pm2$ & $13.\pm3$
   & $14\pm2$ & $11\pm2$\\
\textbf{power law} & & & & & & & & & & \\
   norm$^d$ 
   & $1.07 \pm 0.05$        & $1.60\pm0.16$ & $1.80\pm0.08$ & $1.11\pm0.17$
   & $1.76\pm0.07$ & $1.92\pm0.12$ & $1.85\pm0.05$ & $1.15 \pm 0.02$
   & $1.91\pm0.04$ & $1.49 \pm 0.04$\\
   $\Gamma$
   & $1.23 \pm 0.02$        & $1.42\pm0.05$ & $1.55\pm0.03$ & $1.34 \pm 0.08$
   & $1.42 \pm 0.02$        & $1.46 \pm 0.03$        & $1.44 \pm 0.01$        & $1.24\pm0.02$
   & $1.47 \pm 0.01$        & $1.35 \pm 0.01$\\
\textbf{blackbody} & & & & & & & & & & \\
   norm$^d$ 
   & $3.4\pm0.5$ & $22_{-20}^{+17}$ & $1.20\pm0.13$ & $3.6\pm1.0$
   & $1.8\pm0.4$ & $14.\pm6$   & $1.10\pm0.12$ & $1.95\pm0.25$
   & $1.10\pm0.09$ & $1.88\pm0.21$\\
   $T^c$ 
   & $0.10 \pm 0.01$        & $0.08\pm0.02$ & $0.14 \pm 0.01$      & $0.12\pm0.01$
   & $0.11\pm0.01$ & $0.08\pm0.01$ & $0.12 \pm 0.02$    & $0.11\pm0.01$
   & $0.12 \pm 0.02$ & $0.11 \pm 0.02$\\
\textbf{O VII} & & & & & & & & & & \\
   norm$^e$ 
   & $11 \pm 1$ & $136_{-40}^{+65}$ & $3.3\pm0.9$ & $27_{-7}^{+13}$
   & $9\pm1$ & $40\pm8$ & $3.6\pm0.4$ & $9.6\pm0.8$
   & $2.9\pm0.4$ & $7.6\pm0.7$\\
   $FWHM^f$ 
   & $0.28 \pm 0.04$        & $0.35 \pm 0.06$        & $<0.5$ & $0.22_{-0.06}^{+0.28}$
   & $0.23\pm0.03$ & $0.44\pm0.07$ & $0.28\pm0.05$ & $0.36 \pm 0.05$
   & $0.24\pm0.06$ & $<0.5$\\
 \textbf{O VIII} & & & & & & & & & & \\
   norm$^e$ 
   & $2.7\pm0.3$ & $16_{-4}^{+14}$         & $1.6_{-0.4}^{+1.6}$ & $5._{-2}^{+40}$
   & $2.39_{-0.47}^{+3.74}$ & $8.49_{-2.96}^{+67.96}$ & $0.51\pm0.16$ & $2.32 \pm 0.26$
   & $0.87\pm0.21$ & $1.56\pm0.24$\\
   $FWHM^f$ 
   & $0.16 \pm 0.04$        & $0.34\pm0.09$ & $0.07\pm0.07$ & $<0.11$
   & $0.07\pm0.05$ & $0.05 \pm 0.05$        & $0.20_{-0.08}^{+0.10}$ & $0.10_{-0.02}^{+0.03}$
   & $0.33\pm0.09$ & $0.19\pm0.05$\\

\hline
$\chi^2$/dof & 1.29 & 1.41 & 1.42 & 1.52 & 1.38 & 1.50 & 1.68 & 1.83 & 1.59 & 1.66\\
\hline
\end{tabular}
\end{center}
Notes:\\
$^a$ Column densities ($N_{\rm H}$) are given in units
of $10^{21}$\,cm$^{-2}$ for the neutral, collisionally ionised (CI) and photoionised (PI) absorbers.\\
$^b$ The width $\sigma_v$ of the absorption lines is expressed in \kms.\\
$^c$ Temperatures, $T$, are in eV
for the collisionally ionised gas and in keV for the blackbody.\\
$^d$ The normalisation (norm) of the powerlaw is given in 
$10^{44} \mathrm{\,ph\,s^{-1}\,keV^{-1}}$ at 1\,keV. The normalisation of the blackbody is in units of 
$10^{15} \mathrm{\,cm^{2}}$.\\ 
$^e$ The normalisation of the emission lines (OVII and OVIII) are expressed in units
of  $10^{42} \mathrm{\,ph\,s^{-1}}$.\\
$^f$ The FWHM of the emission lines is in \kms.
\end{table*}

\section{Results}\label{sec: results}

\begin{figure*}
   \centering
   \resizebox{0.45\hsize}{!}{\includegraphics[angle=90]{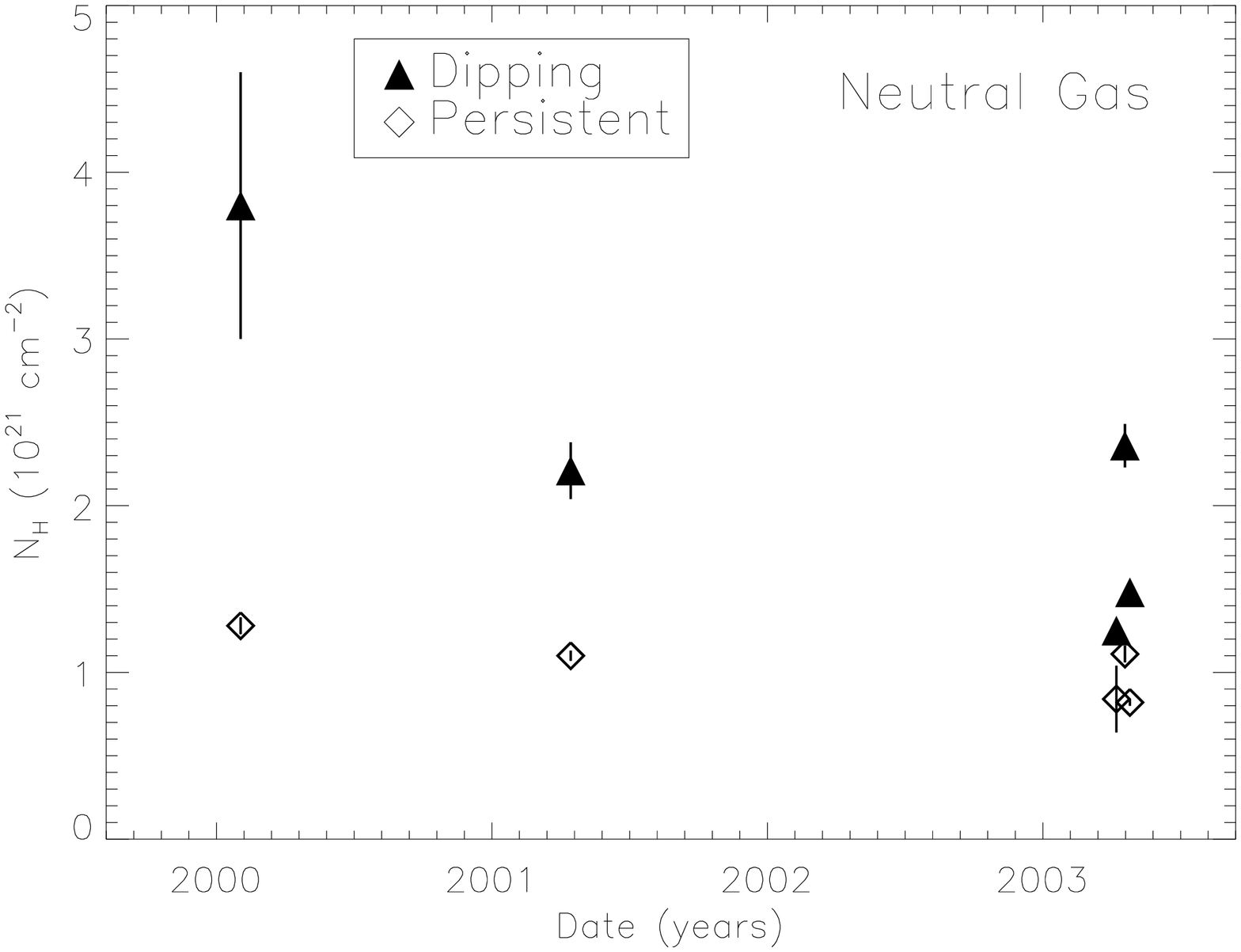}}
   \resizebox{0.45\hsize}{!}{\includegraphics[angle=90]{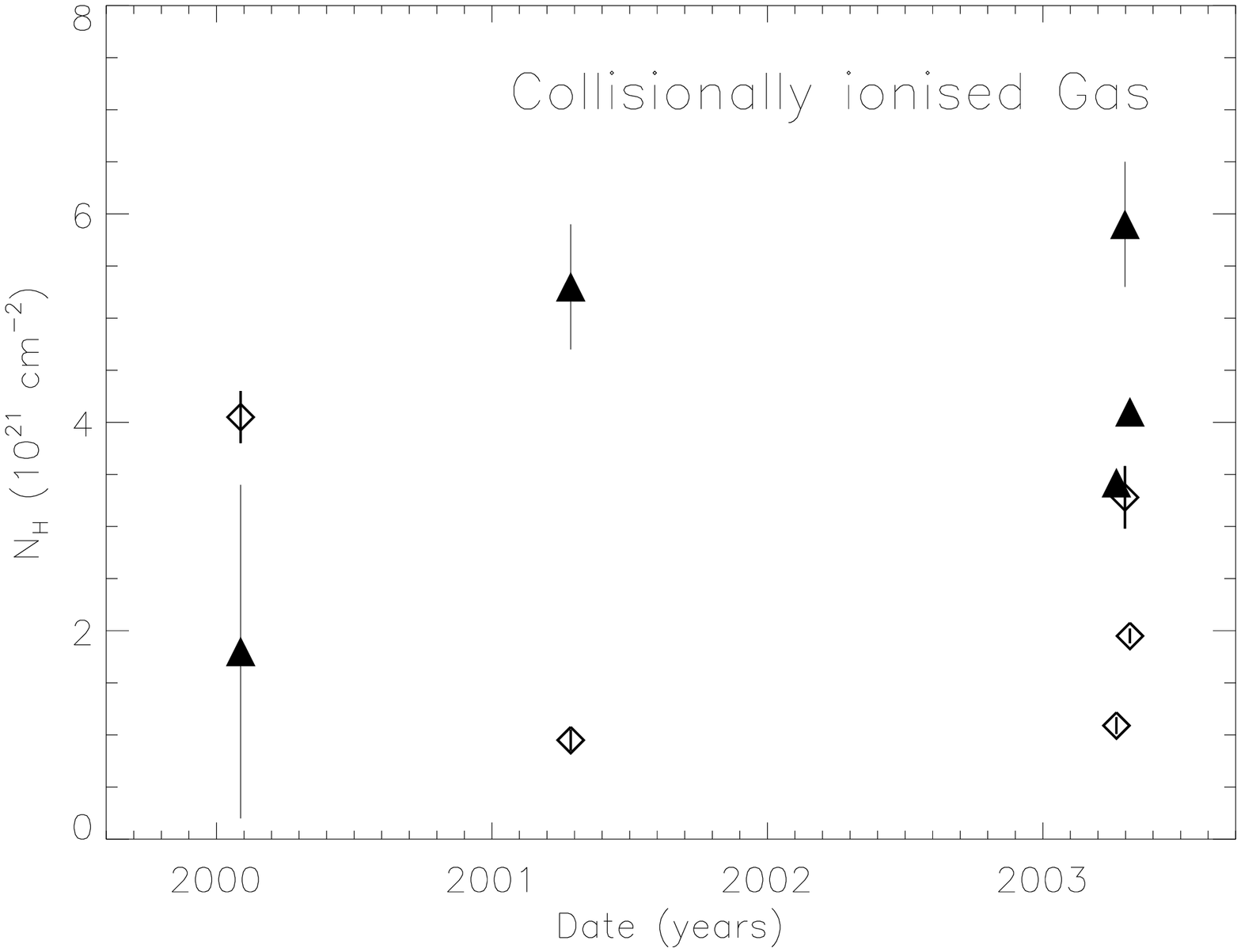}}
   \resizebox{0.45\hsize}{!}{\includegraphics[angle=90]{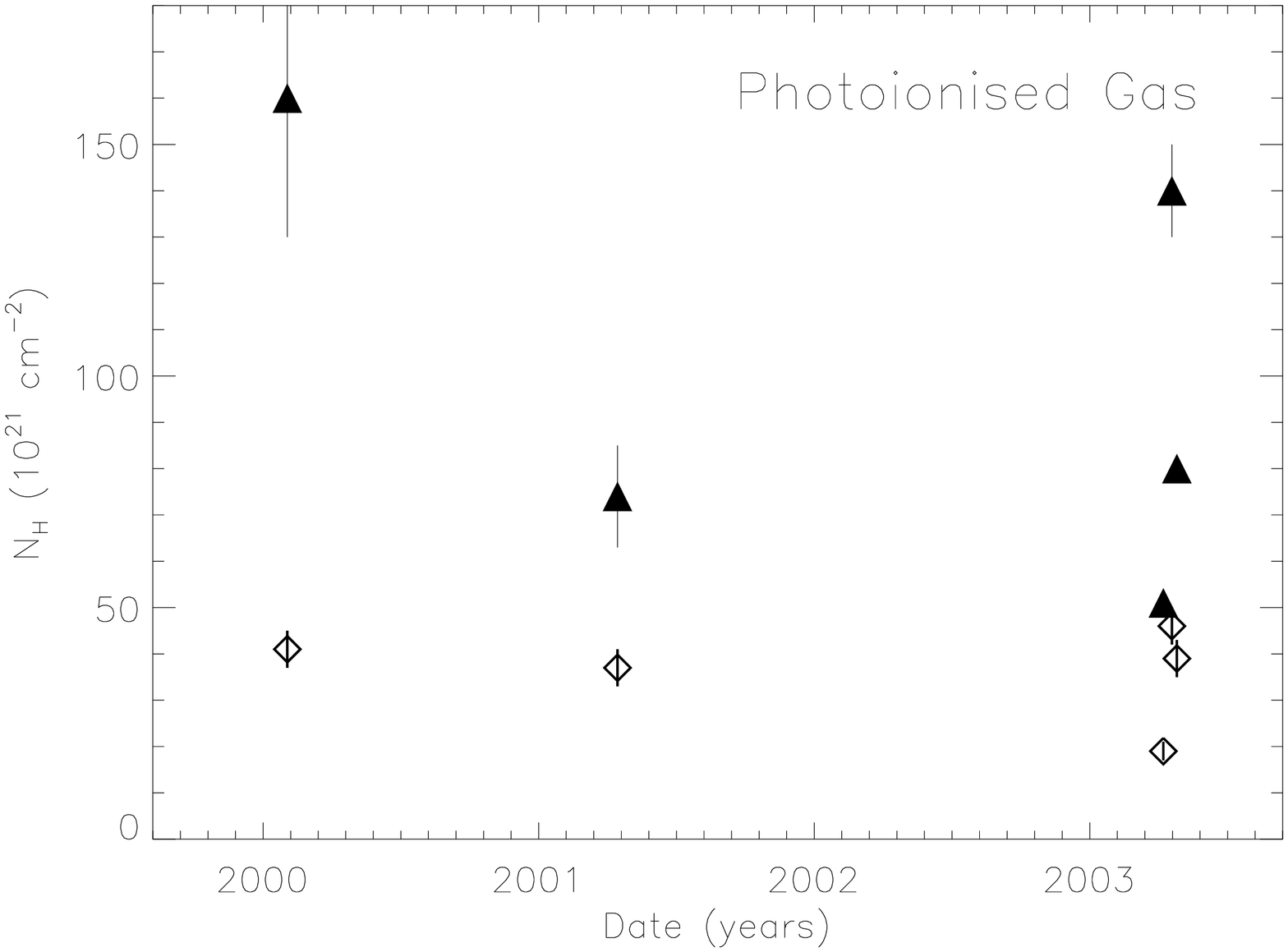}}
   \resizebox{0.45\hsize}{!}{\includegraphics[angle=90]{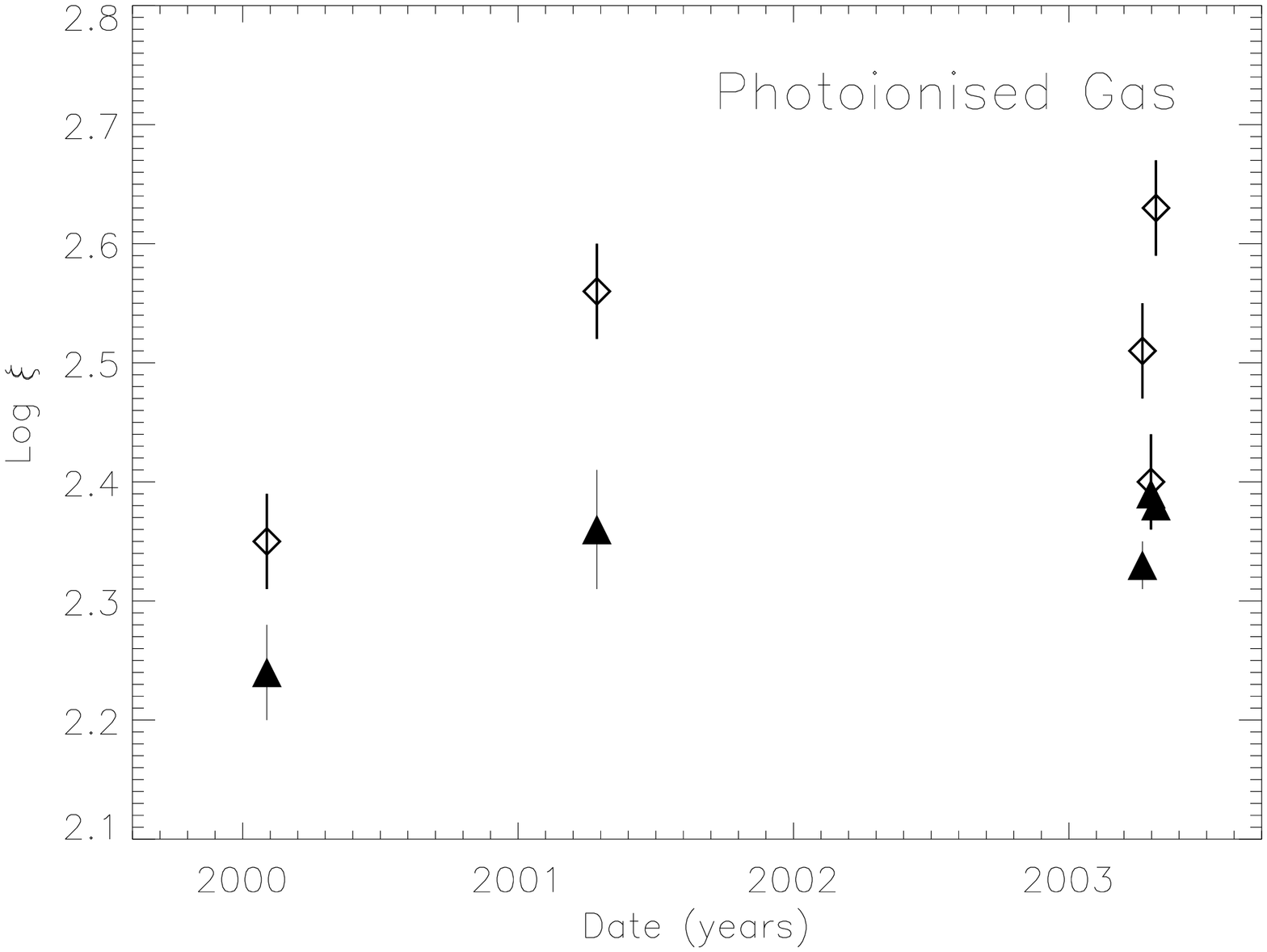}}
   \caption{Parameter values as a function of observation date. 
   Diamonds refer to persistent and triangles to dipping state.
    From top to bottom, left to right:
   neutral absorber column density, collisionally ionised absorber column density,
   photoionised absorber column density and ionisation parameter $\log{\xi}$. Error bars are at $1\sigma$ level.}
   \label{fig: par values}
\end{figure*}

The persistent and the dipping state of \exo\ can be interpreted using the same model for all data groups 
(Table \ref{tab: par values}). While the central source flux changed over time, the continuum shape remained stable. On the
contrary, the absorbers showed significant variability. 
In Fig. \ref{fig: par values} we show the values of the column density of the three absorbers and the ionisation parameter of the
ionised absorber as a function of time. The column density of 
all three absorbers systematically increases during dipping emission.

Fig.~\ref{fig: par values} (first panel) shows that during dips, the column density of the 
neutral gas increases. As we cannot disentangle any intrinsic neutral absorber from the interstellar one, we see
this as an increase of the total neutral absorption. However, the column density in the direction of the system is 
$\sim1.01 \times 10^{21} \mathrm{\,cm^{-2}}$ \citep{2005A&A...440..775K}, 
comparable to the neutral absorber in the persistent emission. This suggests that neutral (or mildly ionised) 
material must exist in the vicinity of the source, in particular near the gas producing the dips.  
 
The collisionally ionised absorber that models mainly the OVII edge 
(Sect.~\ref{sec: results, spectra description}, Fig.~\ref{f:hot_xabs}) 
generally shows an increase in column density during dipping, while the temperature of this component
does not show a clear change. The different behaviour of this ionised absorber in group A
(see Table \ref{tab: par values} and Fig. \ref{fig: par values}) is probably caused by
the low statistics of that group in the dipping spectrum.

The photoionised absorber shows the same behaviour that was already found for this source by \cite{2006A&A...445..179D}: 
during a dip the column density increases, while the ionisation parameter
decreases. Compared to the persistent intervals the absorbing material becomes denser and less
ionised during dips \citep{2005A&A...436..195B}.

For most groups the values of the blackbody normalisation and temperature overlap within errors when we
compare dipping and persistent emission. A moderate scattering is expected due to slow changes of the persistent
emission during long observations. For the power-law parameters the values of $\Gamma$  and normalisation, 
appear instead more scattered when we compare dipping and persistent emission. This may be partially due to cross-calibration effects between
RGS and PN.


\section{Discussion}\label{sec: discussion}

\subsection{Evidence for a collisionally ionised gas}\label{par:ci}
We find that the spectrum of \exo\ can be best modelled by adding a collisionally ionised absorber. 
This model well fits both the persistent and
the dipping emission in all data sets.  
This is the first time that a collisionally ionised absorber is found in a low mass X-ray
binary system. The column density of all the absorbers present in our model increases during dips 
suggesting an association
with the gas producing the dip. A collisionally ionised plasma can exist in the presence of a high ionising radiation field 
($L\sim2\times10^{37}$\,erg\,s$^{-1}$) only if it is located at a sufficiently large distance from the
central source. If we put the gas at such a large distance ($r\sim\ few\times10^{11}$\,cm) and we require
$\xi=L/nr^2<1$, we find $n>few\times10^{14}$\,cm$^{-3}$. 
From the measured column density we then derive a very narrow thickness of the gas $l=N_{\rm H}/n<1.5\times10^{7}$\,cm. 
The geometry of this
gas could then be a uniform shell of gas, with a thickness of less than 150\,km. More realistically, 
the medium could be inhomogeneous, with a low filling factor, as in a smoke-like medium.\\    
In parallel, a further test of the physical process producing the
absorption comes from the associated three OVII emission lines. 
These can be in principle used as a density/temperature
diagnostic \citep[e.g.][]{delphine00,2001A&A...365L.277C}. 
The intercombination ({\it x+y}) 
line is evident in the data while due to absorption some flux from the {\it w} line may be suppressed. 
There is no evidence of the forbidden ({\it z}) OVII line at 22.1\,\AA, suggesting either a high-density
photoionised gas \citep[][]{2001A&A...365L.277C} or a collisionally ionised gas. 
For a photoionised gas we would also expect radiative
recombination continuum of OVII, which is not detected in the data.\\ 
However, both the collisionally ionised and the log$\xi\sim2.5$ 
photoionised plasma included in our model (Table\,\ref{tab: par values}) 
do have a significant amount of OVII 
and OVIII and they can both contribute to the emission spectrum, 
making difficult to disentangle between the two ionising process contributions. The 
measured width of the lines is however large 
(Sect.\,\ref{sec: results, spectra description}). This 
suggests that a considerable amount of matter 
should be located well in the accretion disc, assuming
Keplerian motion \citep[][]{2001A&A...365L.277C}. The FWHM of the detected oxygen lines is roughly 1000\,\kms. Assuming a 1.4$M_{\odot}$
central object then the distance must be of the order of $10^{10}$\,cm.   
Therefore the emission spectrum could be dominated by the 
photoionised gas producing broad lines, 
hiding the narrow emission lines produced by the collisionally ionised gas. 
Indeed, the inferred line width of a collisionally ionised gas rotating at large
($r\sim few\times 10^{11}$\,cm) distance is of the order of 100\,\kms.  
The absence of the OVII
$z$ line is however consistent with our requirement of $n> few\times10^{14}$\,cm$^{-3}$ for the collisionally ionised gas.      
For the photoionised gas, the suppression of the $z$ line could be due either to a high gas density or to a strong UV
radiation field illuminating the gas.

\subsection{The geometry of the system}
Here we test two simple geometries for the absorbing/emitting photoionised gas. We consider the absorption during 
dips as caused by photoionised gas \citep[see][for full details]{2005A&A...436..195B}. 
In both case 1 and 2, described below, we assume that the dipping gas has 
constant  density $n_d$ and it is located at a distance $R_d$ from the central source in a 
protuberance (the impact region) of the disc.\\
As dips are merely caused by absorption, we 
further assume that the intrinsic luminosity of the source remains constant during all the observations 
\citet{2006A&A...445..179D} and \citet{2007A&A...465..559B}.\\ 
Fig.~\ref{fig: geometries} depicts the geometries described in detail below. For both panels, 
the right hand sides describes the observer's view during the dipping state. The left hand sides shows the view during the
persistent emission.

Case 1: the absorption during persistent intervals is due to a thin shell of gas of width $x$ and 
ionisation parameter $\xi_p$ 
(Figure \ref{fig: geometries} and the caption). The shell is
located at a distance $R_{p 1}$. We assume $x$ to be approximately the same as the size 
of the compact impact region that produces the dips (Fig.~\ref{fig: geometries}, upper-right panel). The rationale behind this
is that the gas that produces the persistent absorption could well be gas that trails from the 
impact region into the rest of the disc.
With the assumption that the intrinsic luminosity does not depend on the dipping state of the source, we can compute:

\begin{figure}
\centering
\resizebox{\hsize}{!}{\includegraphics[angle=0]{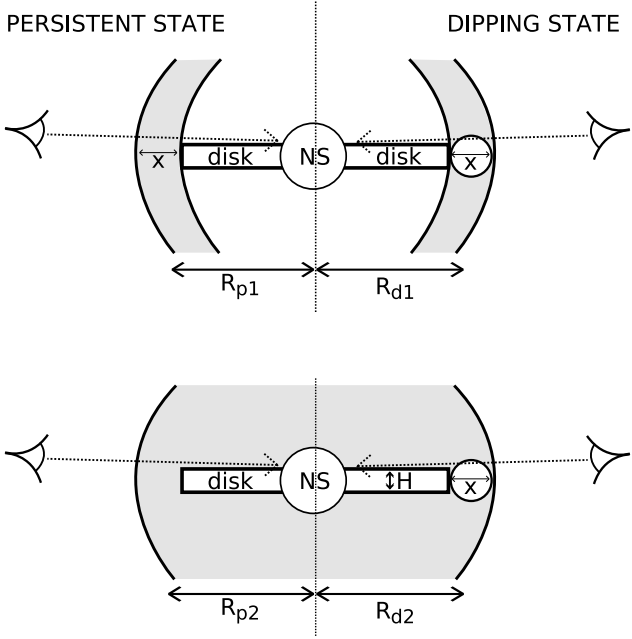}}\\
\caption{Two possible geometries: case 1 in the upper plot the gas producing the persistent absorption has 
a shell-like geometry of thickness $x$ and distance $R_{d1}$ from the central source. In case two (lower plot) the persistent gas 
is a cloud-like geometry of radius $R_{p2}$ encompassing the whole system. 
The labels ``NS'' and ``disc'' indicate the position of the NS and the accretion disc,
respectively. On the right-hand side of the drawings, on both panels,
is the impact region of size $x$ and distance $R_d$ from the central source. The drawn line of sights indicates which
environments the observer intercept during the persistent and dipping state.}
\label{fig: geometries}
\end{figure}

\begin{equation}\label{eq: xid/xip}
\frac{\xi_d}{\xi_p}=\frac{n_p R_{p1}^2}{n_d R_{d1}^2}
\end{equation}
where $\xi$ is the ionisation parameter defined before, $n$ the hydrogen density, $R$ the distance to the ionising
source and the subscripts $p$ and $d$ refer to persistent and dipping emission respectively.

Since we assume that the density has no radial dependence, we can rewrite the above for case 1 as:
\begin{equation}
\frac{R_{p1}}{R_{d1}} = \sqrt{\frac{\xi_d \cdot N_{H,d}}{\xi_p \cdot N_{H,p}}.}
\end{equation}
Here $N_{\rm H}=xn$ is the hydrogen column density derived from our fits for the dipping ($N_{{\rm H},d}$) 
and persistent ($N_{{\rm H},p}$) emission. 
(see \S\ref{sec: results} and table \ref{tab: par values}). \\

Case 2: the persistent gas is in a spherical cloud
encompassing the whole neutron-star-disc system (lower plot in Figure~\ref{fig: geometries}). 
This cloud of photoionised gas with ionisation parameter $\xi_p$ 
has a radius $R_{p2}$  a density $n_p$. The dipping gas is again a 
compact region of width $x$ (Fig.~\ref{fig: geometries}, lower-right panel). 
In this second case we further make the assumption that the disc is geometrically thin \citep{frank, lewin}. 
Therefore, we consider the height of the disc $H\sim0.1R_{d2}$ and also $H\sim x$ (Fig.~\ref{fig: geometries}).\\
The persistent gas is sphere-shaped, therefore $N_{{\rm H},p}=R_{p2}n_p$. For the dipping gas 
$N_{\rm H,d}=xn_d=Hn_d=0.1R_{d2}n_d$, because of the simple relation between $H$ and $R_{d2}$ defined above. 
We can now substitute the values of $N_{\rm H}$ in 
equation (\ref{eq: xid/xip}) and obtain the relation between $R_p$ and $R_d$ for case~2:
\begin{equation}
\frac{R_{p2}}{R_{d2}} = \frac{10 \cdot \xi_d \cdot N_{{\rm H},d}}{\xi_p \cdot N_{{\rm H},p}}
\end{equation}

For both case 1 and case 2, we have calculated the value of $R_p/R_d$ and listed them in Table \ref{tab: geometries}. 
The values for $N_{\rm H}$ and $\xi$ are taken from Tab.~\ref{tab: par values}.

\begin{table}[!htbp]\caption{The ratio of the persistent over dipping radius 
$R_p/R_d$ for the two geometries and for each data group (A--C3).}\label{tab: geometries}
\centering
\begin{tabular}{cccccc}
\hline \hline
 & A & B & C1 & C2 & C3 \\
\hline 
case 1 & $1.7\pm0.4$ & $1.1\pm0.1$ & $1.7\pm0.2$ & $1.1\pm0.1$ & $1.3\pm0.1$ \\
case 2 & $30\pm13$ & $13\pm3$ & $30\pm6$& $11\pm2$ & $17\pm3$ \\
\hline
\end{tabular}
\end{table}
We see that in case 1 the radius of the shell is comparable to the distance of the impact region
to the central source ($R_d \approx R_p$). This is consistent with the idea that the shell could be
a continuation of the impact region. On the contrary, in case 2 the cloud of gas in the persistent state is 
an order of magnitude larger ($R_p \gg R_d$) than the distance of the impact region to the central source. 

\subsection{A flattened geometry for the persistent gas}\label{par:flat}

For both case 1 and 2 described in the previous section, 
it is also possible to put additional constraints on the geometry of the persistent absorbing gas.
This can be achieved using the OVII and OVIII emission lines and the absorption edges
\citep{2001A&A...365L.277C}. 
As justified in Sect.~\ref{par:ci}, we assume here that the emission lines are dominated by photoionisation. 
In this calculation we only use the OVII line and edge for the 
data groups where the associated errors on those values were the
smallest. Therefore we excluded group B. 

We start with the equation for the line luminosity of a given ion $i$ \citep{1999LNP...520..189L}:
\begin{equation}\label{eq: line luminosity}
L^i_{line} = \int n_e^2 \mathrm{d}V \alpha_{RR}(T) \eta_l f_z A \mathrm{~~~[ph/s]},
\end{equation}
where $n_e$ is the electron density, $\alpha_{RR}$ is the radiative recombination coefficient for
the transition, $\eta_l$ is the fraction of recombinations that produce the line, $f_z$ is
the fractional abundance of the ion in ionisation state $i$+1 and $A$ is the absolute abundance
of the element. The radiative recombination coefficient
$\alpha_{RR}$ at 10 eV, which corresponds to the plasma temperature for the best-fit
ionisation parameter, is $\alpha_{RR (OVII)} = 1.6\times10^{-11}
\mathrm{\,cm^3 s^{-1}}$ \citep{1996ApJS..103..467V}.
The fraction of recombinations that produce the OVII line is
$\eta_{l (OVII)}=0.75$ \citep{2001A&A...365L.277C}. The absolute abundance for oxygen is taken from \cite{1989GeCoA..53..197A}:
$A = 8.5114\times10^{-4}$. The emission measure is defined as:
\begin{equation}\label{eq: emission measure}
   EM = \int n_e^2 \mathrm{d}V = \frac{L_{line}}{\alpha_{RR}(T) \eta_l f_z A}
\end{equation}

\noindent
The integral in equation (\ref{eq: emission measure}) can be estimated for both the shell-like (case 1)
and the sphere-like (case 2) geometry for the persistent gas.\\ 
For case 1, for a shell extending from $r$ to $r+x$ ($x\ll r$) we have
\begin{equation}\label{eq: nx}
   n x = \frac{EM_{shell}}{4 \pi n r^2}.
\end{equation}
In fact we consider $r \equiv R_{p}$. For case 2 we simply have for a sphere of gas:
\begin{equation}\label{eq: nr}
   n r =\frac{3}{4\pi}\frac{EM_{sphere}}{n r^2}.
\end{equation}
The value of $n r^2$ can be derived from the ionisation parameter:
\begin{equation}\label{eq: xi}
   n r^2 = \frac{L_{cont}}{\xi}.
\end{equation}
Inserting equation (\ref{eq: xi}), the column densities of equations (\ref{eq: nx}) and (\ref{eq: nr}) 
are now a function of the ionised
absorber, the line luminosity, 
the emission measure and the photoionisation parameter. All these parameters can be evaluated
from the spectral analysis.

\begin{table}
\caption{For each data group, we list the relevant observed and calculated values to obtain the flattening of the gas
(Sect.~\ref{par:flat}). For case 1 and case 2, we also derive the persistent gas column density, 
flattening, density, distance from the central source $R_p$ and its mass. 
We also calculate the distance of the dipping material $R_d$.}
\label{tab: parameter values density and size}
\centering
\begin{tabular}{llllll}
\hline \hline
                                                       & &   A   &  C1  &  C2  &  C3  \\
\hline
{\bf observed values}&&&&&\\
\hline
$L_{cont}$ & $10^{37} \mathrm{\,ergs / s}$         & 1.38  & 1.05  & 1.06  & 0.97 \\
$\log{\xi}$ & $\mathrm{erg\cdot cm}$                     & 2.35  & 2.40  & 2.63  & 2.51 \\
$N_{\rm{H}}^{xabs}$ & $10^{22}$\,cm$^{-2}$& 4.12  & 4.63  & 3.97  & 1.94 \\
\hline
$f_z$               &                                    & 0.317 & 0.274 & 0.132 & 0.200 \\
$L_{line}$ & $10^{42} \mathrm{\,ph / s}$           & 11.2  & 9.04  & 3.63  & 2.97  \\
\hline
{\bf calculated values}&&&&&\\
\hline
$EM$ & $10^{57} \mathrm{\,ph / cm^3}$              & 3.46  & 3.23  & 2.69  & 1.45 \\
$n r^2$ & $10^{34}\mathrm{\,cm^{-1}}$              & 6.16  & 4.18  & 2.48  & 3.00 \\
\hline
\textbf{case 1}&                                         &       &       &       &      \\
$n x$ & $10^{22}\mathrm{\, cm^{-2}}$ & 0.45  & 0.62  & 0.86  & 0.39 \\
$f^{\prime}$=$N_{\rm{H}}^{xabs}/(n x)$ &                   & 9.23  & 7.47  & 4.62  & 4.97 \\

$n$ & $10^{11}\mathrm{\, cm^{-3}}$                 & 27.6  & 50.4  & 64.0 & 12.2 \\

$R_p$ & $10^{11}\mathrm{\, cm}$                    & 1.4  & 0.9  & 0.6  & 1.5 \\
$R_d\ $ & $10^{11}\mathrm{\, cm}$                    & 0.8  & 0.5  & 0.5  & 1.1 \\
$M $ & $10^{-12}\mathrm{\,M_{\odot}}$                & 1.1  & 0.5  & 0.2  & 0.6 \\
\hline
\textbf{case 2} &                                        &       &       &       &      \\
$n r $ & $ 10^{22}\mathrm{\, cm^{-2}}$ & 1.34  & 1.85  & 2.59  & 1.16 \\
$f^{\prime}$=$N_{\rm{H}}^{xabs}/(n r)$ &                   & 3.08  & 2.50  & 1.53  & 1.67 \\

$n $ & $10^{11}\mathrm{\, cm^{-3}}$                 & 0.2  & 0.5  & 0.6  & 0.1 \\

$R_p $ & $10^{11}\mathrm{\, cm}$                    & 14.9  & 9.0  & 6.2  & 15.5 \\
$R_d $ & $10^{11}\mathrm{\, cm}$                    & 0.5 & 0.3  & 0.5  & 0.9 \\
$M $ & $10^{-12}\mathrm{\,M_{\odot}}$                & 22.5  & 8.9  & 3.1  & 10.9 \\
\hline
\end{tabular}
\end{table}

In Table~\ref{tab: parameter values density and size} we compare the expected column density (for case 1, $nx$ and case 2, $nr$) 
of the gas, as derived by the line emission, with the one we measure (from the absorbed spectrum) for each data group.\\ 
Assuming at first that the material subtends a solid angle (as seen from the neutron star) of
$4\pi \mathrm{\,sr}$, the emission of the line will be maximised and, as a consequence, also the derived column density of 
the photoionised emitting gas. 
If we attribute any difference to geometrical effects, 
the ratio between observed/expected column density provides us a flattening factor 
(Table~\ref{tab: parameter values density and size}). From the Table we see that the derived column density 
is systematically
lower than the measured one, suggesting that the 
material does not fully extended in the vertical direction, but rather it is flattened.\\
In case 1 the flattening factor $f^{\prime}$ is $\sim 4.5-9$. 
In case 2 the discrepancy between observed and expected column density is somewhat lower, 
 $\sim 1.5-3$. For case 1, the flattening factor is in agreement with that found previously for this source 
 \citep{2001A&A...365L.277C}.
Converting the flattening factor into an opening angle, 
the ionised persistent gas subtends an average angle of $18^{\circ}$ for case 1 and a wider angle of $54^{\circ}$ for case 2.

\subsection{Density, size, and mass of the photoionised gas}

Using the information we gathered above, 
we calculated the values for the gas density $n$ and distance $R_p$ (Table \ref{tab: parameter values density and size}).
For both geometries we use the identity: $n=(nr)^2/nr^2$.\\
For case 1 we adopt $x\approx 0.1 \times r$ as the thickness of the shell. We obtain:
\begin{eqnarray}\label{eq: n and r case 1}
n & = & 100\frac{(n x f^{\prime})^2}{n r^2} = \left(\frac{10 f^{\prime} L_{line} }{4 \pi \alpha_{RR} \eta f_z A}\right)^2
        \left(\frac{\xi}{L_{cont}}\right)^3,\nonumber\\
r\equiv R_p & = & 0.1\frac{n r^2}{n x f^{\prime}} = \frac{0.4 \pi \alpha_{RR} \eta f_z A}{f^{\prime}
L_{line}}\left(\frac{L_{cont}}{{\xi}}\right)^2.
\end{eqnarray}
For case 2 we have:
\begin{eqnarray}\label{eq: n and r case 2}
n & = & \frac{(n r f^{\prime})^2}{n r^2} = \left(\frac{3 f^{\prime} L_{line} }{4 \pi \alpha_{RR} \eta f_z A}\right)^2
        \left(\frac{\xi}{L_{cont}}\right)^3,\nonumber\\
r\equiv R_p & = & \frac{n r^2}{n r f^{\prime}} = \frac{4 \pi \alpha_{RR} \eta f_z A}{3 f^{\prime} L_{line}}
\left(\frac{L_{cont}}{\xi}\right)^2.
\end{eqnarray}

The average distance from the neutron star to the dipping material $R_d$ is simply obtained 
using the $R_p/R_d$ ratio (Table~\ref{tab: geometries}). 
The distance from the neutron star to
the inner Lagrangian point (L1) varies between $6.4 - 9.7 \times 10^{10} \mathrm{\, cm}$ \citep{frank}.
For both the geometries under study, the location of the dipping material, where the flow of matter from the donor impinges into the accretion
disc, is consistent with the distance of L1. 
The gas seen during the persistent phase is at approximately the same distance of the dipping point in case~1, while in case~2 the sphere of
persistent gas should extend much further. The average thermal velocity of the photoionised gas ($v_{therm}\sim77$\,\kms) 
is more than a factor two lower than the
escape velocity at a distance of $10^{12}$\,cm ($v_{esc}\sim190$\,\kms). 
Therefore in principle the photoionised gas can survive also
beyond L1.\\

Observational evidence of cold gas extending beyond the orbits of the components has 
been found in infra-red for different types of binary systems 
\citep[e.g. cataclysmic variables, novae etc.,][]{deufel99, Solheim94}, including neutron stars 
\citep{2006ApJ...648L.135M}.
The existence of such material, leading possibly to the formation of a circumbinary disc \citep{dubus02}, 
would have important implications for the evolutions of the binary system \citep{spruit01}. 
In this scenario,
the detected collisionally ionised gas (Sect.~\ref{par:ci}) 
could also be a further extension, too far away to be photoionised, 
of the persistent emission. However, the measured temperature of the collisionally 
ionised gas is $\sim$70\,eV, which translates
in a thermal velocity $v_{therm}$ of about 140\,\kms. If the gas is located at a distance
$\gtsim10^{12}$\,cm from the central source $v_{therm}$ becomes comparable to the escape velocity 
$v_{esc}\sim190$\,\kms, 
making the gas a short-lived component, unless continuously replenished by the
accretion disc activity. However, if the distance is $\sim10^{11}$\,cm, the escape velocity already becomes four times
larger than $v_{therm}$, guaranteeing the stability of the collisionally ionised gas. This would favour a 
scenario similar to case~1, where the distance of the persistent and dipping gasses range between 0.5 and 1.4$\times10^{11}$\,cm. 
With this geometry, the collisionally ionised gas could well be an extension of the persistent gas, 
perhaps constituting a first, hotter, 
portion of a circumbinary
disc, which may eventually join a colder phase ($T\sim$600\,K) which extends 
up to four times the distance of the accretion disc \citep[][]{2006ApJ...648L.135M}.\\
In the hypothesis that the collisionally ionised gas is at a 
distance comparable to the persistent and dipping
material, it is likely that there is interaction with the system where the accretion 
stream impacts with the disc, together with
neutral material. A higher concentration (implying a higher column density) 
of the collisionally ionised and neutral gas in correspondence with the dipping
material is then possible \citep[e.g.][and Sect.~\ref{sec: results}]{2005A&A...436..195B}.\\
      
The average gas density is of the order of $few\times 10^{12}$\,cm$^{-3}$ for a shell-like geometry 
(case~1, Table~\ref{tab: parameter values density and size}) and about 100 times less for the spherical geometry 
(case~2, Table~\ref{tab: parameter values density and size}). However, the complete 
suppression of the forbidden line in the persistent spectrum requires a gas density $>10^{14}$\,cm$^{-3}$, unless the gas is irradiated by a 
 strong ultraviolet field. Therefore, if the shape of the OVII triplet 
 is dominated by density effects, we have to require that $n$ is not uniform in the
gas, having for example a radial gradient.\\   
Finally, we calculate the mass of the persistent gas enclosed both in a flattened shell and in a sphere 
using  $ M = n m_p V $
where $M$ is the mass in the gas, $n$ is the density, $m_p$ is the proton mass and $V$ is the
volume of the gas (Table~\ref{tab: parameter values density and size}).


\section{Conclusions}\label{sec: conclusions}
In this paper we present the spectral analysis of a large \xmm\ data set of EXO 0748--676 mainly focusing on the RGS high-energy resolution
data in order to investigate in depth the gaseous environment of the source.\\ 
For both the persistent and dipping spectra, the best fit model is a combination of a neutral absorber,
a collisionally ionised absorber, a photoionised absorber, a power law, a black body and
two Gaussian emission lines at the energies of OVII intercombination line and the OVIII Ly\,$\alpha$. The scatter in the
parameter values for the different data groups gives an idea of the level of uncertainty of our conclusions.\\
Here we summarise the main physical implications of this analysis:

\begin{itemize}

\item The gas during the dipping phase has a higher column density and it is less ionised than the gas in the cloud that permanently 
surrounds the system.\\
This is in line with the idea of ionised absorbers causing the dipping phenomenon
\citep{2005A&A...436..195B, 2006A&A...445..179D}.  In this scenario the compact and colder impact region, where the
accretion stream hits the disc, is in our line of sight during dipping intervals.\\ 

\item We used the information obtained from the RGS spectral analysis to investigate the geometry of 
the persistent/dipping gas and their distance from the
central source.\\ 

\indent First, in the hypothesis of the persistent gas in a shell form, we obtain that 
the thickness and the radius ($\sim few\times10^{11}$\,cm) of the shell are consistent with
those of the dipping gas. The intuitive picture is that the persistent gas is a trailing tail of the dipping gas. 
This axial symmetric geometry should be flattened
with an opening angle of roughly 18$^{\circ}$, consistently with previous studies of this source \citep{2001A&A...365L.277C}.\\ 

\indent In a second scenario, where a spherical cloud of persistent gas was considered, 
we find a smaller flattening, as the angle subtended by the gas is
$\sim$54$^{\circ}$. The persistent gas extends about ten times further than the distance of 
the dipping gas. The radius of the gas sphere is about $0.9-1.5\times10^{12}$\,cm, i.e. larger
than the first Lagrangian point. Photoionised material at such a distance 
might be the first portion of a wider circumbinary disc, detected in infra red in 
at least two neutron stars systems \citep{2006ApJ...648L.135M}.\\ 

For both geometries the gas density is not sufficiently high to 
justify the suppression of the OVII forbidden line in
favour of the OVII intercombination line \citep{delphine00}.\\ 
However, assuming that the emission line is broadened by Keplerian motion around the central
source, we infer a lower limit for the location of the emitting gas of $\sim 10^{10}$\,cm, implying 
a higher gas density ($\sim10^{14}$\,cm$^{-3}$). 
This value is certainly sufficient to produce the observed OVII 
triplet ratios \citep{delphine00}. This suggests that
there may be a radial dependence of the gas density. Another possibility is that the forbidden line in the photoionised
 gas is suppressed by a strong UV field.\\

\item It is the first time that a collisionally ionised absorber is found in a low mass X-ray
binary system. In order for this component to survive to the strong radiation field, it should be located far away from
the central source ($r\sim\ few\times10^{11}$\,cm). This is supported by the evidence that both the temperature and the column
density of such gas change very little during the dipping or persistent phase and seem to be unperturbed on very long
(years) time scale. We estimated that this collisionally ionised absorber should be in the form of dense filaments, with a
very low filling factor. In the hypothesis of a circumbinary disc, this gas might also be an ulterior 
continuation of the persistent gas. In this case the collisionally ionised gas should not be located beyond
$few\times10^{11}$\,cm, in order not to easily escape the system. This would favour a more compact-sized geometry 
for the whole gas environment of the neutron star, as the one
described in case~1, where the persistent gas is a trailing tail of the dipping bulge, at a distance of about $10^{11}$\,cm. 

\end{itemize}

 \begin{acknowledgements}
The authors wish to thank the anonymous referee for his/her useful comments. 
The Space Research Organisation of the Netherlands is financially supported by NWO, The Netherlands Organisation for Scientific Research. 
 \end{acknowledgements}

\bibliographystyle{aa}

\end{document}